\documentclass[prl,floats,aps,twocolumn,epsf,graphicx]{revtex4}
\usepackage{epsfig}
\usepackage{graphicx}

\begin{document}
\title{Near Extreme Black Holes and the Universal Relaxation Bound}
\author{Shahar Hod}
\address{The Ruppin Academic Center, Emeq Hefer 40250, Israel}
\address{ }
\address{The Hadassah Institute, Jerusalem 91010, Israel}
\date{\today}

\begin{abstract}

\ \ \ A fundamental bound on the relaxation time $\tau$ of a
perturbed thermodynamical system has recently been derived, $\tau
\geq \hbar/\pi T$, where $T$ is the system's temperature. We
demonstrate {\it analytically} that black holes saturate this bound
in the extremal limit and for large values of the azimuthal number
$m$ of the perturbation field.

\end{abstract}
\bigskip
\maketitle


The deep connections between the physics of black holes, and the
realms of thermodynamics and information theory were revealed by
Hawking's theoretical discovery of black-hole radiation \cite{Haw},
and its corresponding black-hole entropy and temperature
\cite{Beken1}. These discoveries suggest that black holes behave as
thermodynamic systems in many respects.

On another front, a fundamental problem in thermodynamic and
statistical physics is to determine the relaxation timescale at
which a perturbed physical system returns to a stationary,
equilibrium configuration \cite{Kubo}. Recently, we have derived
\cite{qeb} a universal bound on the relaxation time $\tau$ of
perturbed thermodynamical systems:

\begin{equation}\label{Eq1}
\tau_{min} = \hbar/ \pi T\  ,
\end{equation}
where $T$ is the system's temperature \cite{Units}. This bound is
based on standard results from quantum information theory and
thermodynamic considerations \cite{Bekinf,Shanon}. Thus the
relaxation time of a perturbed physical system is fundamentally
bounded by the reciprocal of its temperature: the colder it gets the
longer it takes to settle down to a stationary equilibrium
configuration, in accord with the spirit of the third-law of
thermodynamics \cite{NoteNam}.

It should be noted that mundane physical systems are characterized
by relaxation times which are typically many orders of magnitude
larger than $\hbar /T$ \cite{qeb}. To put our relaxation bound to an
interesting test, we must consider physical systems with strong
self-gravity.

Back to Gravity: a stationary black hole corresponds to a thermal
state, characterized by the Bekenstein-Hawking temperature
\cite{Haw,Beken1}, $T_{BH}={{\hbar (r_+ -r_-)} \over {4\pi(r^2_+
+a^2)}}$, where $r_{\pm}=M+(M^2-a^2)^{1/2}$ are the black-hole outer
and inner horizons, and $M$ and $a$ are the black-hole mass and
angular momentum per unit mass, respectively. (We consider here the
canonical Kerr black holes). Perturbing the black hole corresponds
to perturbing this thermal state, and the decay of the perturbation
(characterized by damped oscillations) describes the return to
thermal equilibrium.

The response of a black hole to external perturbations is
characterized by `quasinormal ringing' (QNM), damped oscillations
with a discrete frequency spectrum \cite{Nollert1}. All
perturbations of the black-hole spacetime are radiated away in a
manner reminiscent of the last pure dying tones of a ringing bell:
the relaxation process is characterized by an exponential decay of
the form $e^{-i\omega t}$, with complex black-hole quasinormal
frequencies, $\omega =\Re\omega -i\Im\omega$.

It turns out that a perturbed black hole has an infinite number of
quasinormal frequencies, characterizing oscillations with decreasing
relaxation times (increasing imaginary part) \cite{Leaver}. The mode
with the smallest imaginary part (known as the fundamental mode)
gives the dynamical timescale $\tau$ for generic perturbations to
decay (the relaxation time required for the perturbed black hole to
return to a quiescent state). Namely, $\tau \equiv \omega^{-1}_I$,
where $\omega_I$ denotes the imaginary part of the fundamental,
least damped black-hole resonance.

Taking cognizance of the relaxation bound Eq. (\ref{Eq1}), one
deduces an upper bound on the black-hole fundamental frequency
\cite{qeb}

\begin{equation}\label{Eq2}
\omega_I \leq \pi T_{BH}/\hbar \  .
\end{equation}
Thus the relaxation bound implies that a black hole must have (at
least) one quasinormal resonance whose imaginary part conform to the
upper bound (\ref{Eq2}). This mode would dominate the relaxation
dynamics of the perturbed black hole, and will determine its
characteristic relaxation timescale.

It has been demonstrated numerically \cite{qeb} that perturbed black
holes conform to the relaxation bound Eq. (\ref{Eq2}). In fact,
black holes have relaxation times which are of the same order of
magnitude as $\tau_{min}$, the minimally allowed relaxation time
\cite{qeb}. Here we shall prove {\it analytically} that black holes
may actually attain the fundamental relaxation bound in the extremal
limit.

Mashhoon \cite{Mash} has calculated the QNM spectrum of rotating
black holes in the eikonal limit $l=|m| \gg 1$, where $m$ is the
azimuthal number of the perturbation field. The Kerr QNM frequencies
in the $l=m \gg 1$ limit are given by \cite{Mash}

\begin{equation}\label{Eq3}
\omega_n =m\omega_+  -i(n+{1\over 2})\beta\omega_+\ \ \ ; \ \ \
n=0,1,2,...\ ,
\end{equation}
where

\begin{equation}\label{Eq4}
\omega_+\equiv {{M^{1/2} \over{r^{3/2}_{ph}+aM^{1/2}}}}\ ,
\end{equation}
is the Kepler frequency for null rays in the unstable equatorial
circular orbit of the black hole, and

\begin{equation}\label{Eq5}
r_{ph}\equiv 2M\{1+\cos[{2 \over 3}\cos^{-1}(-a/M)]\}\  ,
\end{equation}
is the limiting circular photon orbit. The function $\beta$ is given
by \cite{Mash}

\begin{equation}\label{Eq6}
\beta= {{(12M)^{1/2}(r_{ph}-r_+)(r_{ph}-r_-)} \over
{r^{3/2}_{ph}(r_{ph}-M)}}
\end{equation}

We now focus on the extremal limit, $T_{BH} \to 0$. Let
$r_{\pm}=M\pm\epsilon$, where $\epsilon \ll 1$ [this implies
$T_{BH}={\epsilon/{4\pi M^2}}+O(\epsilon^2)$]. After some algebra we
find that Eq. (\ref{Eq5}) yields

\begin{equation}\label{Eq7}
r_{ph}=M+{2\epsilon \over {\sqrt{3}}} +O(\epsilon^2)\  ,
\end{equation}
which in turn implies

\begin{equation}\label{Eq8}
\omega_+ ={1 \over {2M}}-{{\sqrt{3} \epsilon}\over
{4M^2}}+O(\epsilon^2)\ ,
\end{equation}
and

\begin{equation}\label{Eq9}
\beta={\epsilon \over M} +O(\epsilon^2).
\end{equation}
Substituting Eqs. (\ref{Eq8}) and (\ref{Eq9}) in Eq. (\ref{Eq3}), we
find that the near-extreme black-hole quasinormal resonances are
given by

\begin{equation}\label{Eq10}
\hbar\omega_n =m\hbar\Omega -i2\pi T_{BH} \cdot (n+{1\over 2})\ ,
\end{equation}
where $\Omega \equiv {a \over {r^2_+ +a^2}}=1/2M+O(\epsilon)$ is the
angular velocity of the black-hole event horizon.

It has long been known qualitatively that the QNM tend to cluster
near the real axis in the extreme black-hole limit \cite{Det}.
However, the quantitative description of the imaginary part which is
the focus of the present analysis [that is, $\Im(\omega_n)=2\pi
T_{BH}(n+{1 \over 2})$ in the $T_{BH} \to 0$ limit] is a new result
and was not established explicitly in previous studies.

Finally, taking $n=0$ in Eq. (\ref{Eq10}), one finds

\begin{equation}\label{Eq11}
{{\hbar\omega_I} \over {\pi T_{BH}}} \to 1^{-}\  ,
\end{equation}
for the fundamental, least damped black-hole resonance
\cite{noteneg}. This implies that Kerr black holes have the
remarkable property of saturating the universal relaxation bound in
the extremal limit. In this limit, the black hole has the shortest
possible relaxation time for a given temperature.

\bigskip

\noindent {\bf ACKNOWLEDGMENTS}
\bigskip

This research is supported by the Meltzer Science Foundation. I
thank Andrei Gruzinov for interesting correspondence and Uri Keshet
for stimulating discussions.


\begin{thebibliography}{99}
\bibitem{Units} We use gravitational units in which $G=c=k_B=1$.
\bibitem{noteneg} It is straightforward to verify that the
correction term of order $O(T^2_{BH})$ in Eq. (\ref{Eq10}) has a
negative imaginary part, and thus the approach to $2\pi T_{BH}(n+{1
\over 2})$ in the extremal limit is from below.
\end{thebibliography}
\end{document}